\documentclass[10pt, conference]{IEEEtran}

\usepackage{graphicx}
\usepackage{subfigure,tikz}
\usepackage{amssymb}
\usepackage[utf8]{inputenc}
\usepackage{bm,bbm}
\usepackage{booktabs}
\usepackage{multirow}
\usepackage{rotating}
\usepackage{microtype}

\usepackage[cmex10]{amsmath}
\usepackage{amsthm}

\usepackage{hyperref}

\begin{document}

\title{SAR Image Despeckling Algorithms using Stochastic Distances and Nonlocal Means}
\newif\iffinal
\finaltrue
\newcommand{\jemsid}{114723}

\thanks{Dissertation presented at \textit{Programa de P\'os-Gradua\c c\~ao em Modelagem Computacional de Conhecimento}, Universidade Federal de Alagoas, Macei\'o, AL}

  \author{%
    \IEEEauthorblockN{%
      Leonardo Torres and Alejandro C.\ Frery
      }
    \IEEEauthorblockA{%
      Laborat\'orio de Computa\c c\~ao Cient\'ifica e An\'alise Num\'erica -- LaCCAN\\
      Universidade Federal de Alagoas -- UFAL\\
      57072-970, Macei\'o, AL -- Brazil\\
      Email: \{ljmtorres, acfrery\}@gmail.com
    }
  }


\maketitle

\begin{abstract}
This paper presents two approaches for filter design based on stochastic distances for intensity speckle reduction.
A window is defined around each pixel, overlapping samples are compared and only those which pass a goodness-of-fit test are used to compute the filtered value.
The tests stem from stochastic divergences within the Information Theory framework.
The technique is applied to intensity Synthetic Aperture Radar (SAR) data with homogeneous regions using the Gamma model.
The first approach uses a Nagao-Matsuyama-type procedure for setting the overlapping samples, and the second uses the nonlocal method.
The proposals are compared with the Improved Sigma filter and with anisotropic diffusion for speckled data (SRAD) using a protocol based on Monte Carlo simulation.
Among the criteria used to quantify the quality of filters, we employ the equivalent number of looks, and line and edge preservation.
Moreover, we also assessed the filters by the Universal Image Quality Index and by the Pearson correlation between edges.
Applications to real images are also discussed.
The proposed methods show good results.
\end{abstract}


\IEEEpeerreviewmaketitle

\section{Introduction}\label{sec:intro}

Synthetic Aperture Radar (SAR) data are generated by a system of coherent illumination and are affected by the interference coherent of the signal.
It is known that these data incorporate a granular noise, known as speckle noise,  that degrades its quality.
This noise is also present in the laser, ultrasound-B, and sonar imagery~\cite{Goodman1976}.

The speckle phenomenon in SAR data hinders the interpretation of these data and reduces the accuracy of segmentation, classification and analysis of objects (targets) contained within the image.
Therefore, reducing the noise effect is an important task, and multilook processing is often used for this purpose in single-channel data.

Lee et al.~\cite{Lee1999PolSARspeckleFiltering,Lee1991MultiPol_SARimagery} proposed techniques for speckle reduction based on the multiplicative noise model using the minimum mean-square error (MMSE) criterion.
Lee et al.~\cite{Lee2006ScatteringModelBased_SpeckleFiltering} proposed a methodology for selecting neighboring pixels with similar scattering characteristics, known as Refined Lee filter.
The Improved Sigma filter~\cite{Lee2009} is an improvement of the previous proposals, where an undesired blurring was solved by redefining the sigma range based on the speckle probability density functions.

{\c C}etin and Karl~\cite{Cetin2001FeatureEnhancedSAR} presented a technique for image filtering based on regularized image reconstruction.
This approach employs a tomographic model which allows the incorporation of prior information about, among other features, the sensor.
The resulting images have many desirable properties, reduced speckled among them.
Our approach deals with data already produced and, thus, does not require interfering in the processing protocol of the data.

Osher et al.~\cite{Osher2005IteratedRegularizationMethod} presented a novel iterative regularization method for inverse problems based on the use of Bregman distances using a total variation denoising technique tailored to additive noise.
The authors also propose a generalization for multiplicative noise, but no results with this kind of contamination are shown. 
The main contributions were the rigorous convergence results and effective stopping criteria for the general procedure, that provides information on how to obtain an approximation of the noise-free image intensity.

More recent proposals are based on nonlocal (NL) means method originally proposed by Buades et al.~\cite{Buades2005DenoisingAlgorithms} which is based on the redundancy of neighboring patches.
The noise-free estimated value of a pixel is defined as a weighted mean of pixels in a certain region.
Under the Additive White Gaussian Noise (AWGN) assumption, these weights are calculated using Euclidean distances to measure the similarity between a central region patch and neighboring patches in a search window.
However, the speckle noise is not well described by a Gaussian distribution requiring, thus, changes in the model.

Deledalle et al.~\cite{NonGaussianPatchSimilarity} analyzed several similarity criteria for data which depart from the Gaussian assumption, viz., the Gamma and Poisson noises.
In~\cite{Deledalle2009IterativeDenoising} the same authors extended the NL-means method to speckled imagery using statistical inference in an iterative procedure.
The authors derived the weights using the likelihood function of Gaussian and square root of Gamma (termed ``Nakagami-Rayleigh'') noises.
In~\cite{Deledalle2010NonlocalSARandInSAR}, the authors proposed the use of a nonlocal approach to estimate jointly reflectivity, phase difference and coherence from a pair of co-registered single-look complex SAR images.

Coup\'e et al.~\cite{Coupe2009NLM_FilteringUltrasound} also used a logarithmic transformation and assume zero-mean Gaussian noise to propose the Optimized Bayesian NL-means with block selection (OBNLM).
The OBNLM filter is an optimized version of the filter proposed by Kervrann et al.~\cite{Kervrann2007BayesianNLM} which employs a new distance for comparing patches and then selecting the most similar pixels.

Parrilli et al.~\cite{Parrilli2012} presented a nonlocal technique based on Block-Matching 3D for SAR images (SAR-BM3D) inspired by the algorithm presented in~\cite{Dabov2007} for AWGN denoising.
The SAR-BM3D filter has two steps: first, the algorithm estimates the noise-free image, and the second step, the algorithm filters anew using the more reliable statistics computed on the basic estimate to improve the filter performance.

Statistical analysis is essential for dealing with speckled data.
Different statistical distributions are proposed in the literature to describe speckle data.
It provides comprehensive support for developing procedures for interpreting the data efficiently, and to simulate plausible images.
In this paper we use the Gamma distribution to describe the speckle noise, and a constant to characterize the ground truth~\cite{Gao2010}.

This paper presents two approaches for speckle noise filtering: the first, a local nonlinear procedure, and the second, an adaptive nonlinear extension of the NL-means algorithm introduced by Buades et al.~\cite{Buades2005DenoisingAlgorithms}.
The first approach~\cite{Torres2012IGARSS,SpeckleReductionStochasticDistancesCIARP2012}, 
termed Stochastic Distances Nagao-Matsuyama (SDNM) filter, uses the neighborhoods defined by Nagao and Matsuyama~\cite{NagaoMatsuyama} around each pixel; samples are compared and only those which pass a goodness-of-fit test based on stochastic distances between distributions.
The test is based on a stochastic distance whose good statistical properties stem from the Information Theory framework.
The improvement of previous works, the second approach, we called Stochastic Distance Nonlocal Means (SDNLM).
An improvement of this latest proposal applied to PolSAR data is found in~\cite{TorresSpeckleReductionPolSAR}.

The paper is organized as follows:
Section~\ref{sec:model} presents the statistical modeling used to describe speckle data.
In Section~\ref{sec:distances} samples are compared and only those which pass a goodness-of-fit test based on stochastic distances between distributions.
Section~\ref{sec:assessment} presents the metrics for assessing the quality of the filtered images.
Sections~\ref{sec:results} and~\ref{sec:conclu} present the results and conclusions, respectively.

\section{The Multiplicative Model}\label{sec:model}

According to~\cite{Goodman1976}, the multiplicative model can be used to describe SAR data.
This model asserts that the intensity observed in each pixel is the outcome of the random variable $Z$ which, in turn, is the product of two independent non-negative random variables: $X$, that characterizes the mean radar reflectivity or radar cross section; and $Y$, which models speckle noise.
The law which describes the observed intensity $Z=XY$ is completely specified by the distributions proposed for $X$ and $Y$.

This paper assumes locally homogeneous intensity images, so the constant scale parameter $X=\lambda>0$ defines the backscatter, and the unitary-mean Gamma distribution models the multilook speckle noise.
Thus, it follows that $Z\sim \Gamma\left(L,{L}/{\lambda}\right)$ and its density is
\begin{equation}\label{eq:densGamma}
 f_Z(z;L,\lambda) = \frac{L^L}{\lambda^L\Gamma(L)} z^{L-1} \exp\Big\{ -\frac{Lz}{\lambda} \Big\},\quad z>0,
\end{equation}
where $\Gamma$ stands for the Gamma function and $L\geq 1$ is the equivalent number of looks.
We describe different levels of heterogeneity by allowing the number of looks $L$ to vary locally.
In a similar way, S{\o}lbo \& Eltoft~\cite{WMAPSpeckleFilterWavelet} assume a Gamma distribution in a wavelet-based speckle reduction procedure, and they locally estimate all the parameters without imposing a fixed number of looks (which they call ``degree of heterogeneity'') for the whole image. 
These authors use a large $33\times 33$ neighborhood to estimate this parameter, whereas we employ small windows.

The likelihood of $\bm{z}=(z_1,z_2,\dots,z_n)$, a random sample of size $n$ from the $\text{Gamma}(L,L/\lambda)$ law, is given by
\begin{equation}
\mathcal{L}(L,\lambda;\bm{z})=\bigg( \frac{L^L}{\lambda^L\Gamma(L)} \bigg)^n \prod_{j=1}^{n} z_j^{L-1} \exp\Big\{ -\frac{Lz_j}{\lambda} \Big\}.
\end{equation}
Thus, the maximum likelihood estimator for $(L,\lambda)$, namely $(\widehat{L},\widehat{\lambda})$, is given by $\widehat\lambda = n^{-1}\sum_{j=1}^n z_j$ and by the solution of 
\begin{equation}
 \ln\widehat{L} - \psi^0(\widehat{L}) - \ln \frac1n \sum_{j=1}^n z_j + \frac1n \sum_{j=1}^{n} \ln z_j = 0,
\end{equation}
 where $\psi^0$ is the digamma function~\cite{SpeckleReductionStochasticDistancesCIARP2012}.

\section{Stochastic Distances Filter}\label{sec:distances}

\subsection{Neighborhood of the first approach}

The first filter, initially proposed in~\cite{Torres2012IGARSS,SpeckleReductionStochasticDistancesCIARP2012}, is local and nonlinear.
It is based on stochastic distances and tests between distributions~\cite{Nascimento2010}, obtained from the class of ($h,\phi$)-divergences.
The proposal employs the neighborhoods defined by Nagao and Matsuyama~\cite{NagaoMatsuyama}.

Each filtered pixel has a $5\times5$ neighborhood, within which nine areas are defined and treated as different samples.
Denote $\bm{\widehat{\theta}_1}$ the estimated parameter in the central $3\times3$ neighborhood,
and $\big(\bm{\widehat{\theta}}_2,\ldots,\bm{\widehat{\theta}}_{9}\big)$ the estimated parameters in the eight remaining areas.
To account for possible departures from the homogeneous model, we estimate $\bm{\widehat{\theta}}_i=(L_i,\lambda_i)$, $i=\{1,\dots,9\}$ by maximum likelihood.
The filtered value is the result of averaging the central patch with those that pass the goodness-of-fit test at a certain level of confidence stipulated by the user.

\subsection{Neighborhood of the second approach}

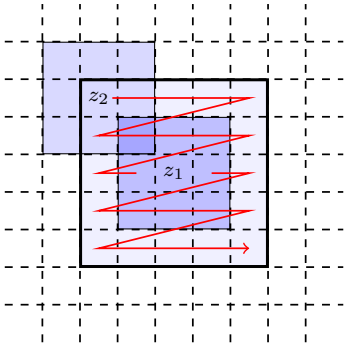
\begin{figure}[hbt]
\begin{center}
\begin{tikzpicture}
\node () at (1.75,1.75)[fill=blue!06,shape=rectangle,draw=black,text width=2.25cm,text height=2.25cm]{};
\node () at (1.75,1.75)[fill=blue!25,shape=rectangle,draw=black,text width=1.25cm,text height=1.25cm]{};
\node () at (1.75,1.75)[]{\footnotesize$z_1$};

\node () at (0.75,2.75)[fill=blue!15,shape=rectangle,draw=black,text width=1.25cm,text height=1.25cm]{};
\node () at (0.75,2.75)[]{\footnotesize$z_2$};

\node () at (1.25,2.25)[fill=blue!35,shape=rectangle,draw=black,text width=0.25cm,text height=0.25cm]{};
\node () at (1.75,1.75)[shape=rectangle,draw=black,line width=.4mm,text width=2.25cm,text height=2.25cm]{};

  \draw[very thick,dashed,line width=.23mm] (0.0,-0.5) -- (0.0,4.0);
  \draw[very thick,dashed,line width=.23mm] (0.5,-0.5) -- (0.5,4.0);
  \draw[very thick,dashed,line width=.23mm] (1.0,-0.5) -- (1.0,4.0);
  \draw[very thick,dashed,line width=.23mm] (1.5,-0.5) -- (1.5,4.0);
  \draw[very thick,dashed,line width=.23mm] (2.0,-0.5) -- (2.0,4.0);
  \draw[very thick,dashed,line width=.23mm] (2.5,-0.5) -- (2.5,4.0);
  \draw[very thick,dashed,line width=.23mm] (3.0,-0.5) -- (3.0,4.0);
  \draw[very thick,dashed,line width=.23mm] (3.5,-0.5) -- (3.5,4.0);
  
  \draw[very thick,dashed,line width=.23mm] (-0.5,0.0) -- (4.0,0.0);
  \draw[very thick,dashed,line width=.23mm] (-0.5,0.5) -- (4.0,0.5);
  \draw[very thick,dashed,line width=.23mm] (-0.5,1.0) -- (4.0,1.0);
  \draw[very thick,dashed,line width=.23mm] (-0.5,1.5) -- (4.0,1.5);
  \draw[very thick,dashed,line width=.23mm] (-0.5,2.0) -- (4.0,2.0);
  \draw[very thick,dashed,line width=.23mm] (-0.5,2.5) -- (4.0,2.5);
  \draw[very thick,dashed,line width=.23mm] (-0.5,3.0) -- (4.0,3.0);
  \draw[very thick,dashed,line width=.23mm] (-0.5,3.5) -- (4.0,3.5);
  
  \draw[-, red,very thick,line width=.23mm] (0.93,2.75) -- (2.75,2.75) -- (0.75,2.25) -- (2.75,2.25) -- (0.75,1.75) -- (1.25,1.75);
  \draw[->,red,very thick,line width=.23mm] (2.25,1.75) -- (2.75,1.75) -- (0.75,1.25) -- (2.75,1.25) -- (0.75,0.75) -- (2.75,0.75);
\end{tikzpicture} 
\end{center}
\vspace*{-.5cm}
\caption{Central pixel $z_1$ and its neighboring $z_i$, $i=\{2,\dots,25\}$ with $3\times3$ pixels patches.}\label{fig:patches}
\end{figure}

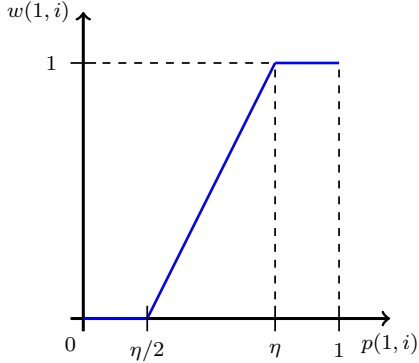
\begin{figure}[hbt]
\begin{center}
\begin{tikzpicture}[scale=1.7]
\node () at (-0.1,-0.2)[]{\footnotesize$0$};
\node () at (-0.25,2.0)[]{\footnotesize$1$};
\node () at (-0.35,2.4)[]{\footnotesize$w(1,i)$};
\node () at (0.5,-0.25)[]{\footnotesize$\eta/2$};
\node () at (1.5,-0.25)[]{\footnotesize$\eta$};
\node () at (2.0,-0.25)[]{\footnotesize$1$};
\node () at (2.4,-0.20)[]{\footnotesize$p(1,i)$};
  \draw[->,very thick,line width=.4mm] (-0.1,0) -- (2.4,0);
  \draw[->,very thick,line width=.4mm] (0,-0.1) -- (0,2.4);
  \draw[blue,very thick,line width=.35mm] (0,0) -- (.5,0);
  \draw[blue,very thick,line width=.35mm] (.5,0) -- (1.5,2);
  \draw[very thick,line width=.23mm] (-0.1,2) -- (0.1,2);
  \draw[very thick,line width=.23mm] (0.5,-0.1) -- (0.5,0.1);
  
  \draw[blue,very thick,line width=.35mm] (1.5,2) -- (2,2);
  \draw[very thick,line width=.23mm] (1.5,-0.1) -- (1.5,0.1);
  \draw[very thick,dashed,line width=.23mm] (1.5,0.15) -- (1.5,2);
  
  \draw[very thick,line width=.23mm] (2,-0.1) -- (2,0.1);
  \draw[very thick,dashed,line width=.23mm] (2,0.15) -- (2,2);
  \draw[very thick,dashed,line width=.23mm] (0.15,2) -- (1.5,2);
\end{tikzpicture} 
\end{center}
\vspace*{-.5cm}
\caption{Weight function for every pair of patches $(1,i)$, $2\leq i\leq 25$.}\label{fig:WeightFunction}
\end{figure}

In this approach the neighborhoods of the central pixel and of its surrounding pixels are of the same size: $3\times3$ pixels. 
The central patch, with center pixel $z_1$, is thus compared with $24$ neighboring patches, whose center pixels are $z_i$, $i=2,\dots,25$, as illustrated in Figure~\ref{fig:patches}.
The estimate of the noise-free observation at $z_1$ is a weighted sum of the observations at $z_2,\dots,z_{25}$, being each weight a function of the $p$-value ($p({1,i})$) observed in the statistical test of same distribution between two Gamma laws: 
\begin{equation}
 w(1,i) = 
\left\{
\begin{array}{ll} 
1  &\text{if }\; p({1,i}) \geq \eta, \\ 
\frac2\eta p({1,i})-1 & \text{if }\; \frac{\eta}{2} < p({1,i}) < \eta, \\
0 & \text{otherwise},
\end{array}
\right.
\label{eq:weightNLMeansTorres}
\end{equation}
where $\eta$ is the confidence of the test, chosen by the user.
This function is illustrated in Figure~\ref{fig:WeightFunction}.
In this way we employ a soft threshold instead of an accept-reject decision.
This allows the use of more evidence than with a binary decision.

In this way, this proposal employs those pixels whose observations are not rejected by a test of strong isotropy with respect to the central value.

\subsection{Stochastic Distances Filter}

The proposals are based on the use of stochastic distances on small areas within the filtering window.
Consider $Z_1$ and $Z_i$ random variables defined on the same probability space, characterized by the densities $f_{Z_1}(z_1;\bm{\theta}_1)$ and $f_{Z_i}(z_i;\bm{\theta}_i)$, respectively, where $\bm{\theta_1}$ and $\bm{\theta_i}$ are parameters.
Assuming that both densities have the same support $I \subset \mathbbm{R}$, the ($h,\phi$)-divergence between the distributions is given by
\begin{equation}
D_{\phi}^{h}(Z_1,Z_i) = h \Big( \int_{x\in I}\;\phi \Big( \frac{f_{Z_1}(x;\bm{\theta}_1)}{f_{Z_i}(x;\bm{\theta}_i)} \Big) \;f_{Z_i}(x;\bm{\theta}_i)\;\mathrm{d}x \Big),
\end{equation}
where $h\colon (0,\infty)\rightarrow[0,\infty)$ is a strictly increasing function with $h(0)=0$ and $h'(x)>0$; and $\phi\colon (0,\infty)\rightarrow[0,\infty)$ is a convex function for all $x \in \mathbbm{R}$.
Choices of the functions $h$ and $\phi$ result in several divergences.

Divergences sometimes do not obey the requirements to be considered distances.
A simple solution is to define a new measure, the distance $d_{\phi}^{h}$, given by
%
\begin{equation}
d_{\phi}^{h}(Z_1,Z_i) = \frac{D_{\phi}^{h}(Z_1,Z_i)+D_{\phi}^{h}(Z_i,Z_1)}{2}.
\end{equation}
%
Distances, in turn, can be conveniently scaled in order to present good statistical properties that make them test statistics~\cite{Nascimento2010}:
\begin{equation}
S_{\phi}^{h}(\bm{\widehat{\theta}}_1,\bm{\widehat{\theta}}_i)=\frac{2mnk}{m+n}\;d^{h}_{\phi}(\bm{\widehat{\theta}}_1,\bm{\widehat{\theta}}_i),
\end{equation}
where $\bm{\widehat{\theta}}_1$ e $\bm{\widehat{\theta}}_i$ are maximum likelihood estimators based on samples size $m$ and $n$, respectively, and $k=\big(h'(0)\phi''(1)\big) ^{-1}$.
The null hypothesis $\bm{\theta_1}=\bm{\theta_i}$ is rejected at a level $\eta$, if $\Pr(S_{\phi}^{h} > s)\leq\eta)$, where $s$ is the observed value.
Under mild conditions $S_{\phi}^{h}$ is $\chi^2_M$ asymptotically distributed, being $M$ the dimension of $\bm{\theta_1}$, the test is well defined.
Details can be seen in the work by Salicr\'u et al.~\cite{Salicru1994}. 
Several statistical tests were derived (Hellinger, Bhattacharyya, Triangular, $\chi^2$, and R\'enyi of order $\beta$), and the one with the best computational performance was the one based on the Kullback-Leibler divergence:
\begin{equation}
S_{K\!L} = \frac{mn\big(\widehat{L}_1+\widehat{L}_i\big)}{m+n}\;\bigg(\frac{\widehat{\lambda}_1^2+\widehat{\lambda}_i^2}{2\widehat{\lambda}_1 \widehat{\lambda}_i}-1\bigg).
\end{equation}

The filtering procedure consists in checking which regions can be considered as coming from the same distribution that produced the data which comprises the central block.
The sets which are not rejected are used to compute a local mean.
If all the sets are rejected, the filtered value is updated with the average on the $3\times3$ neighborhood around the filtered pixel.

\section{Image Quality Assessment}\label{sec:assessment}

\begin{figure}[hbt] %
 \centering
 \subfigure[Phantom\label{fig:phantom}]{\includegraphics[width=.325\linewidth]{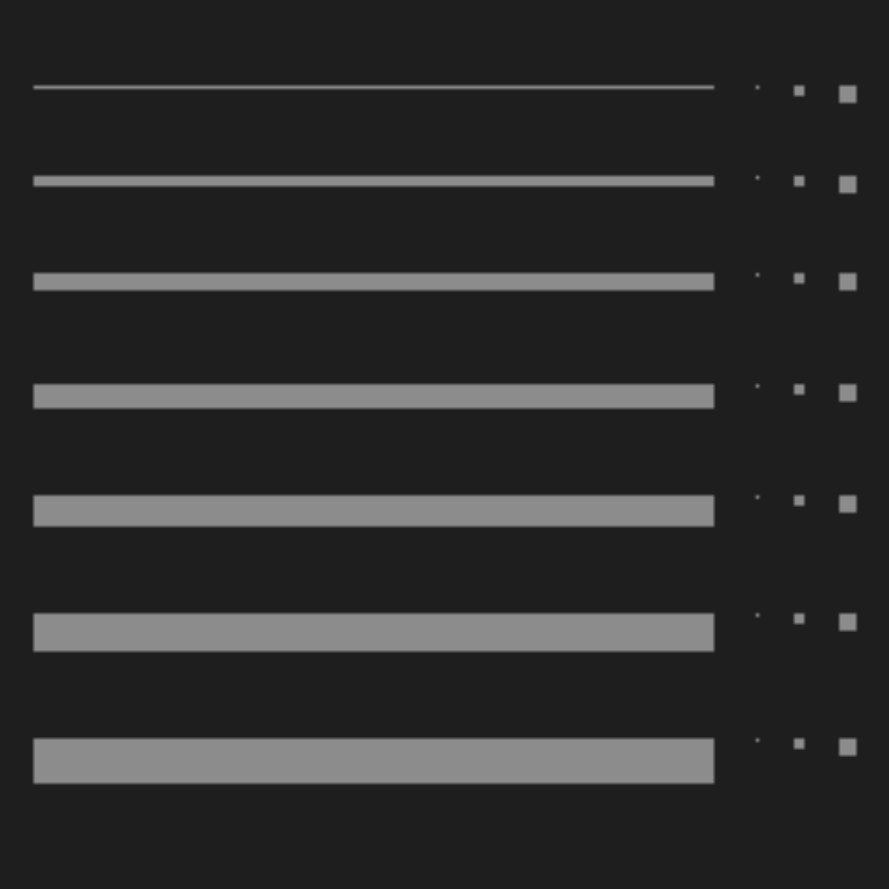}}
 \subfigure[Corrupted, $3$-looks\label{fig:corrupt}]{\includegraphics[width=.325\linewidth]{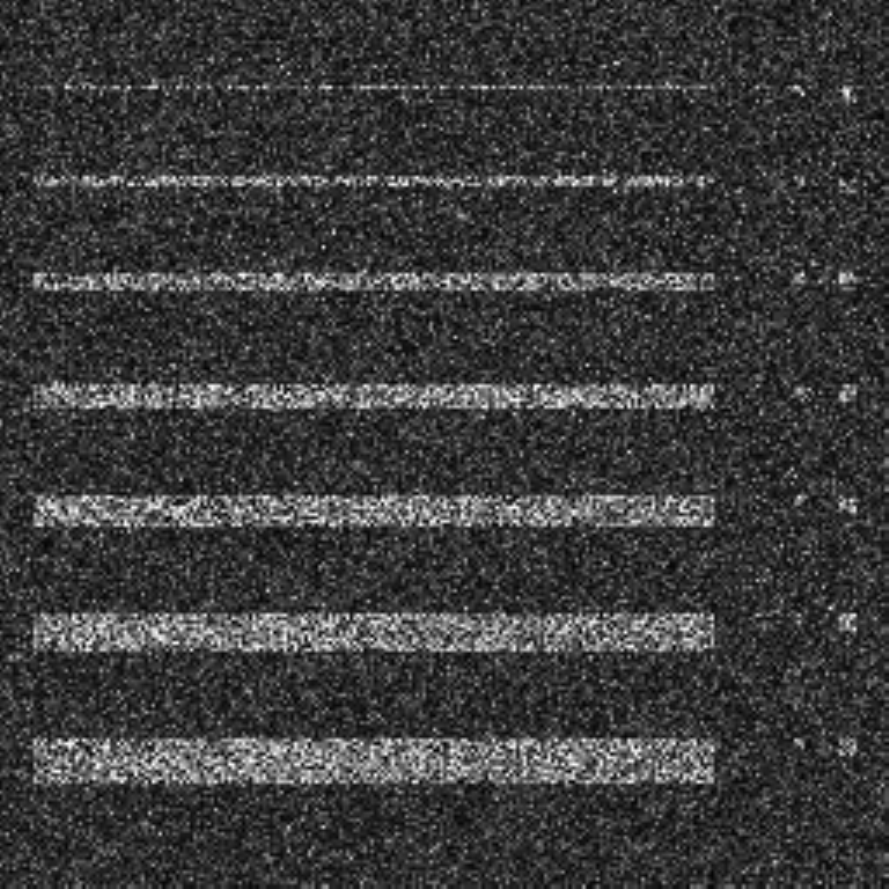}}
 \subfigure[Improved Sigma filter\label{fig:LeeFilter}]{\includegraphics[width=.325\linewidth]{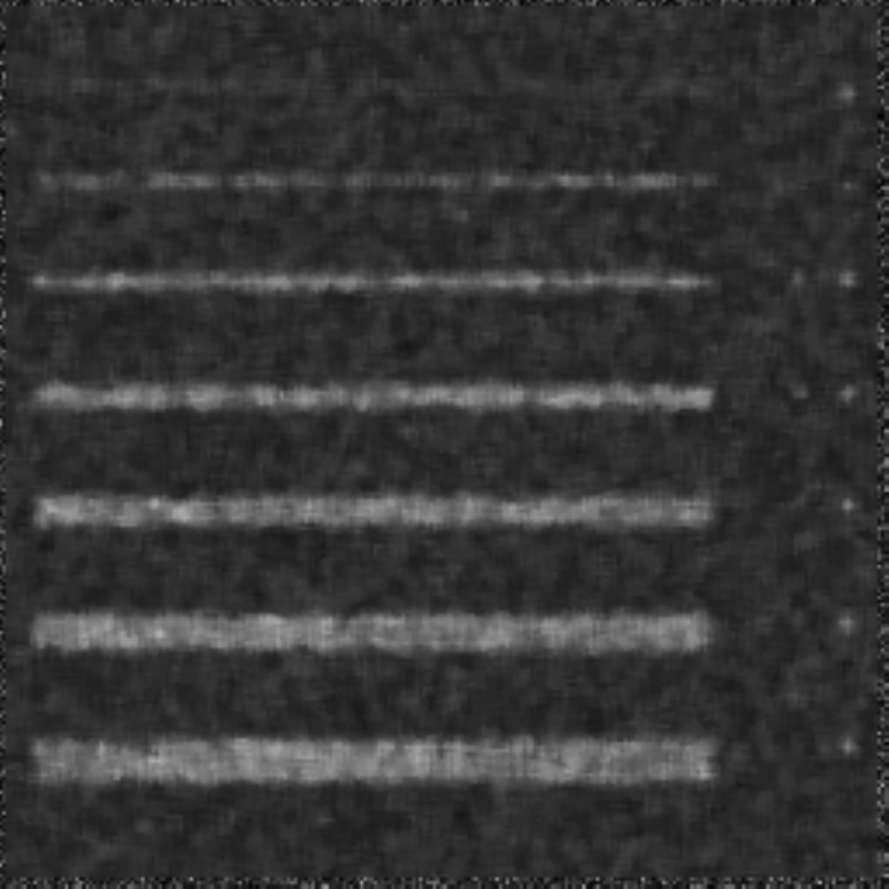}}
 \subfigure[SRAD filter\label{fig:img_SRAD}]{\includegraphics[width=.325\linewidth]{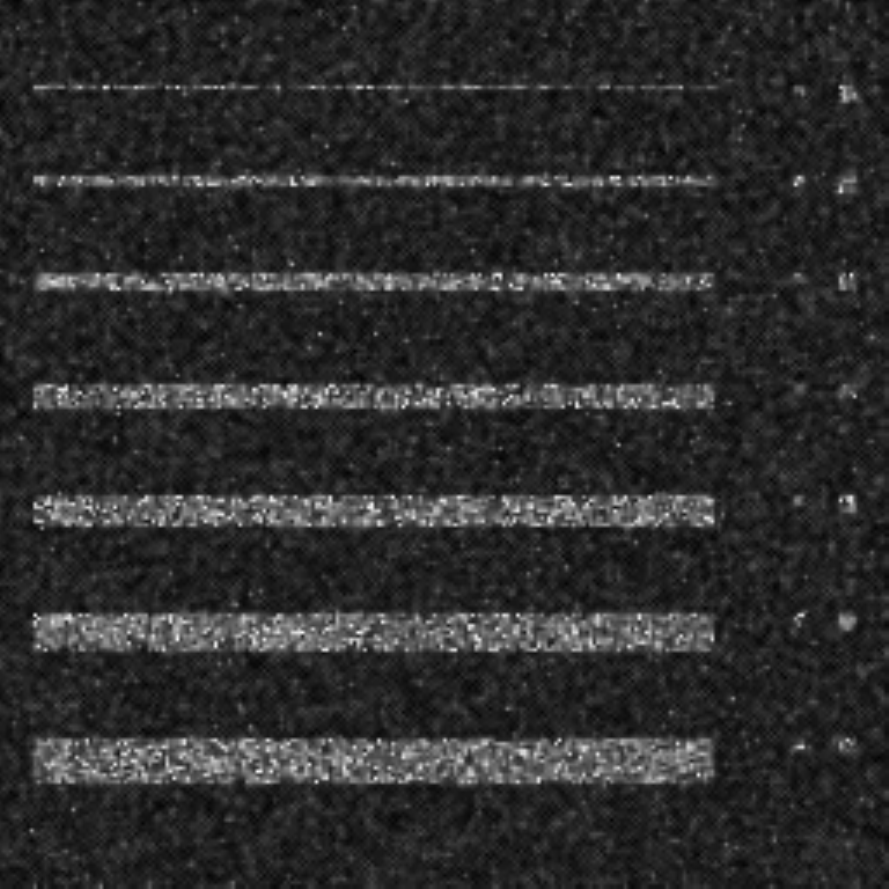}}
 \subfigure[SDNM filter\label{fig:KL_Filter}]{\includegraphics[width=.325\linewidth]{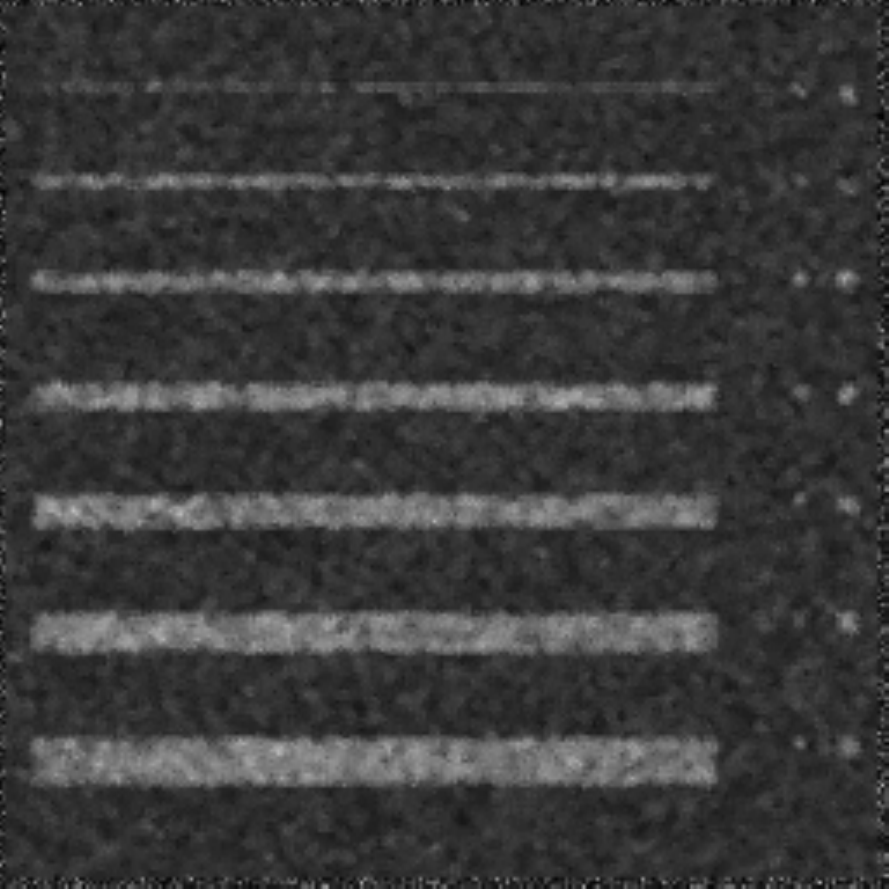}}
 \subfigure[SDNLM filter\label{fig:NL_Filter}]{\includegraphics[width=.325\linewidth]{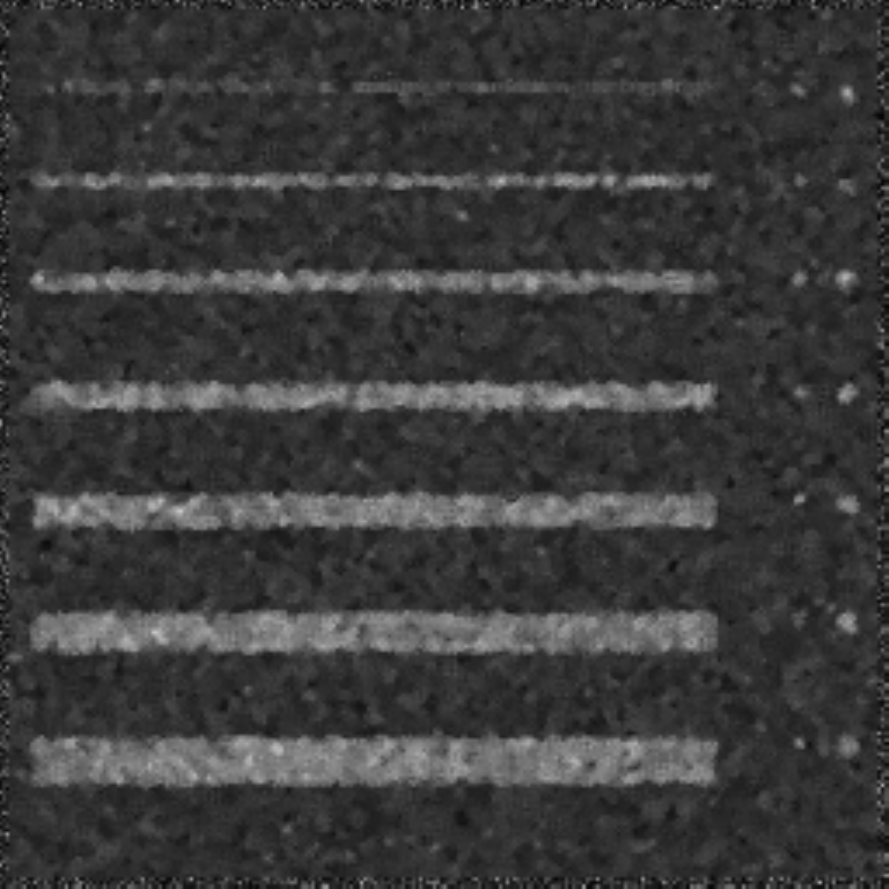}}
 \caption{Lee’s Protocol phantom, speckled data and filtered images.}
 \label{fig:protocol}
\end{figure}

Image quality assessment in general, and filter performance evaluation in particular, are hard tasks~\cite{UIQIndex}.
Moschetti et al~\cite{Moschetti2006} discussed the need of making a Monte Carlo study when assessing the performance of image filters.
They proposed a protocol which consists of using a phantom image (see Figure~\ref{fig:phantom}) corrupted by speckle noise (see Figure~\ref{fig:corrupt}). 
The experiment consists of simulating corrupted images as matrices of independent samples of some distribution with different parameters.
Every simulated image is subjected to filters, and the results are compared (see Figures~\ref{fig:LeeFilter} to~\ref{fig:NL_Filter}).

Among the criteria used to quantify the quality of the filters, we employ~\cite{Moschetti2006}: 
\begin{itemize}
 \item \textbf{Equivalent Number of Looks}: in intensity imagery and homogeneous areas, it can be estimated by $\textsf{ENL}=(\bar{z}/\widehat{\sigma}_Z)^2$, i.e., the square of the reciprocal of the coefficient of variation. In this case, the bigger the better.
 
 \item \textbf{Line Contrast}: the pre\-serva\-tion of a line of one pixel of width will be assessed by computing three means: in the coordinates of the original line ($x_{\ell}$) and in two lines around it ($x_{\ell_1}$ and $x_{\ell_2}$). The contrast is then defined as $2x_{\ell}-(x_{\ell_1} + x_{\ell_2})$, and compared with the contrast in the phantom. The best values are the smallest.
 
 \item \textbf{Edge Preserving}: it is measured by means of the edge gradient (the absolute difference of the means of strip around edges) and variance (same as the former but using variances instead of means). The best values are the smallest.
\end{itemize}
A ``good'' technique must combat speckle and, at the same time, preserve details as well as relevant information.

Furthermore, we also assessed the filters by the universal image quality index \cite{UIQIndex}, the correlation measure $\beta_{\rho}$ and the $\operatorname{BRISQUE}$ model \cite{Mittal2012NoReferenceIQASpatialDomain} on real images.
The universal image quality index is defined by
\begin{equation}
Q = \frac{s_{xy}}{s_x s_y} \frac{2\overline{xy}}{\overline{x}^2 + \overline{y}^2} \frac{2 s_x s_y}{s_x^2 + s_y^2},
\end{equation}
where $s_\bullet^2$ and $\overline{\bullet}$ denote the sample variance and mean, respectively.
The range of $Q$ is $[-1,1]$, being $1$ the best value.
The quantity
\begin{equation}
\beta_{\rho} = \frac{\sum_{j=1}^{n} (x_j-\bar{x})(y_j-\bar{y})}{\sqrt{\sum_{j=1}^{n} (x_j-\bar{x})^2 \sum_{j=1}^{n} (y_j-\bar{y})^2}},
\end{equation}
is a correlation measure is between the Laplacians of images $X$ and $Y$, where $\bullet_j$ and $\overline{\bullet}$ denote the gradient values of the $j$th pixel and mean of the images $\nabla^2 X$ and $\nabla^2 Y$, respectively. 
The range of $\beta_{\rho}$ is $[-1,1]$, being $1$ perfect correlation.

The $\operatorname{BRISQUE}$ is a model that operates in the spatial domain and requires no-reference image.
This image quality evaluator does not compute specific distortions such as ringing, blurring, blocking, or aliasing, but quantifies possible losses of ``naturalness'' in the image. 
This approach is based on the principle that natural images possess certain regular statistical properties that are measurably modified by the presence of distortions. No transformation to another coordinate frame (DFT, DCT, wavelets, etc) is required, distinguishing it from previous blind/no-reference approaches.
The $\operatorname{BRISQUE}$ is defined for scalar-valued images and it ranges in the $[0,100]$ interval, and smaller values indicate better results.

\begin{figure*}[ht]
 \centering
 \includegraphics[width=.95\linewidth]{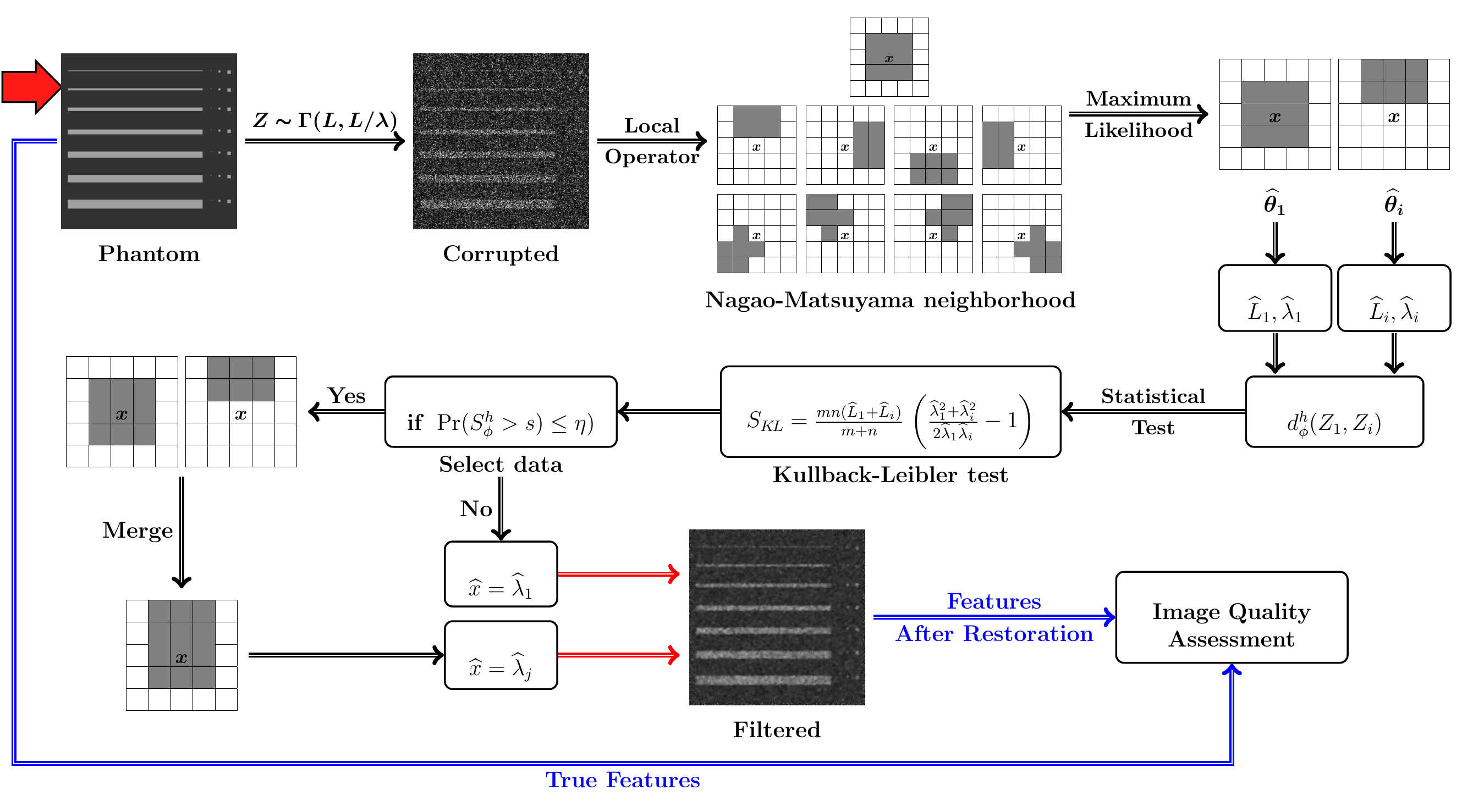}
 \caption{Block diagram for assessment of the first proposed technique.}
 \label{fig:diagrama_blocos}
\end{figure*}

Figure~\ref{fig:diagrama_blocos} shows a block diagram of the assessment method for the first proposal.

\section{Results and Analysis}\label{sec:results}

The proposals were compared with the Improved Sigma filter~\cite{Lee2009} and the SRAD (Speckle Reduction Anisotropic Diffusion) filter proposed by Yu and Acton~\cite{Yu2002SpeckleReducingAnisotropicDiffusion}, specifically designed for combating speckle.
The Improved Sigma filter was applied in windows of sides $5$.
The SRAD filter used a window of side $5$, and diffusion threshold $q_0=1/2$.
The tests of the stochastic distances filters were performed at the $90\%$ level of significance.

Table~\ref{tab:simulated_data_I} presents the three situations that were simulated.
These parameters describe situations commonly found when analyzing SAR imagery in homogeneous regions.

\begin{table}[hbt]
 \centering
 \caption{Simulated situations with the $\text{Gamma}(L,L/\lambda)$ distribution.}\label{tab:simulated_data_I}
 \begin{tabular}{c c c c} \toprule
  Situation ID & ~$L$~ & ~$\lambda_{\ell}$~ & Background mean \\ \midrule
       $\#1$     &  $1$  &  $200$      & $20$            \\ 
       $\#2$     &  $3$  &  $195$      & $55$            \\
       $\#3$     &  $4$  &  $150$      & $30$            \\ \bottomrule
 \end{tabular}
\end{table}

The results obtained in one hundred independent replications are summarized in Table~\ref{tab:statistics}: the mean and the standard deviation (in parenthesis) of each measure are shown in each situation.
Only the results of applying the filter once are presented, for $L=\{1,3,4\}$ looks following the Table~\ref{tab:simulated_data_I}, and the best results are highlighted in bold.

\begin{table*}[hbt]
\setlength{\tabcolsep}{1pt}
\centering
\caption{Statistics of the metrics of simulated images: $100$ replications with $1$-iteration.}\label{tab:statistics}
\begin{tabular}{l r@{.}l r@{.}l r@{.}l r@{.}l r@{.}l r@{.}l r@{.}l r@{.}l r@{.}l r@{.}l r@{.}l r@{.}l} 
\toprule
\textbf{Filtered} & \multicolumn{16}{c}{\textbf{SAR Measures}} & \multicolumn{4}{c}{\multirow{2}{*}{\textbf{$Q$ Index}}} & \multicolumn{4}{c}{\multirow{2}{*}{\textbf{$\beta_{\rho}$ Index}}} \\ \cmidrule(lr{.5em}){2-17}
\textbf{Versions} & \multicolumn{4}{c}{\textsf{ENL}} & \multicolumn{4}{c}{Line Cont.} & \multicolumn{4}{c}{Edge Grad.} & \multicolumn{4}{c}{Edge Var.} & \multicolumn{4}{c}{~} & \multicolumn{4}{c}{~} \\ 
\midrule
Improved Sigma~$\#1$ & \textbf{14}&\textbf{375} & (2&034) & 1&724 & (0&032) & 74&806 & (10&065)& \textbf{2}&\textbf{117} & (1&426) & 0&149 & (0&002) & 0&747 & (0&005) \\ 
SRAD~$\#1$           &  1&009 & (0&144) & 1&798 & (0&027) & 84&217 & (8&066) & 2&533 & (1&108) & 0&001 & (0&000) & 0&793 & (0&007) \\ 
SDNM~$\#1$           & 13&391 & (1&333) & 1&566 & (0&035) & \textbf{61}&\textbf{975} & (6&900) & 4&533 & (1&234) & 0&220 & (0&002) & 0&812 & (0&009) \\  
SDNLM~$\#1$          & 12&054 & (3&393) & \textbf{1}&\textbf{522} & (0&041) & 64&531 & (6&789) & 5&020 & (1&256) & \textbf{0}&\textbf{226} & (0&002) & \textbf{0}&\textbf{822} & (0&008) \\ 
\midrule
Improved Sigma~$\#2$ & \textbf{54}&\textbf{467} & (7&226) & 1&495 & (0&029) & 62&709 & (5&038) & 6&772 & (1&492) & 0&198 & (0&002) & 0&784 & (0&006) \\ 
SRAD~$\#2$           &  3&158 & (0&421) & 1&639 & (0&042) & 72&925 & (5&583) & 6&444 & (1&250) & 0&001 & (0&000) & 0&811 & (0&011) \\ 
SDNM~$\#2$           & 40&234 & (3&978) & 1&393 & (0&041) & \textbf{61}&\textbf{975} & (6&900) & 6&736 & (1&321) & 0&235 & (0&002) & 0&840 & (0&010) \\  
SDNLM~$\#2$          & 43&495 & (6&514) & \textbf{1}&\textbf{361} & (0&060) & 64&531 & (6&789) & \textbf{5}&\textbf{899} & (1&644) & \textbf{0}&\textbf{243} & (0&001) & \textbf{0}&\textbf{845} & (0&010) \\ 
\midrule
Improved Sigma~$\#3$ & \textbf{90}&\textbf{238} & (10&939)& 1&401 & (0&024) & 47&230 & (6&051) & 8&315 & (1&728) & 0&218 & (0&001) & 0&845 & (0&002) \\ 
SRAD~$\#3$           &  4&765 & (0&850) & 1&511 & (0&046) & 63&068 & (8&751) & \textbf{3}&\textbf{528} & (0&875) & 0&001 & (0&000) & 0&863 & (0&005) \\ 
SDNM~$\#3$           & 65&678 & (7&451) & 1&293 & (0&037) & \textbf{37}&\textbf{255} & (4&928) & 6&053 & (1&619) & 0&248 & (0&001) & 0&883 & (0&003) \\  
SDNLM~$\#3$          & 66&485 & (19&580)& \textbf{1}&\textbf{207} & (0&063) & 47&866 & (3&581) & 4&765 & (1&518) & \textbf{0}&\textbf{262} & (0&001) & \textbf{0}&\textbf{899} & (0&007)  \\ 
\bottomrule
\end{tabular} 
\end{table*}

As expected, the Improved Sigma filter (denoted as Lee) is the one which provides the strongest speckle reduction as measured by the equivalent number of looks (\textsf{ENL}).
This filter was designed with that purpose in mind.
When it comes to measures of detail preservation, our proposal is the winner.


Not every aforementioned quality measure can be applied to real data, unless the ground truth is known.
One of the quality measures that can be used in this case is the $\operatorname{BRISQUE}$ index~\cite{Mittal2012NoReferenceIQASpatialDomain}.

\begin{figure*}[hbt]
\centering
  \subfigure[SAR data (HH polarization)\label{fig:SARdataFoulum}]{\includegraphics[viewport= 351 1 600 350,clip=TRUE,width=.18\linewidth]{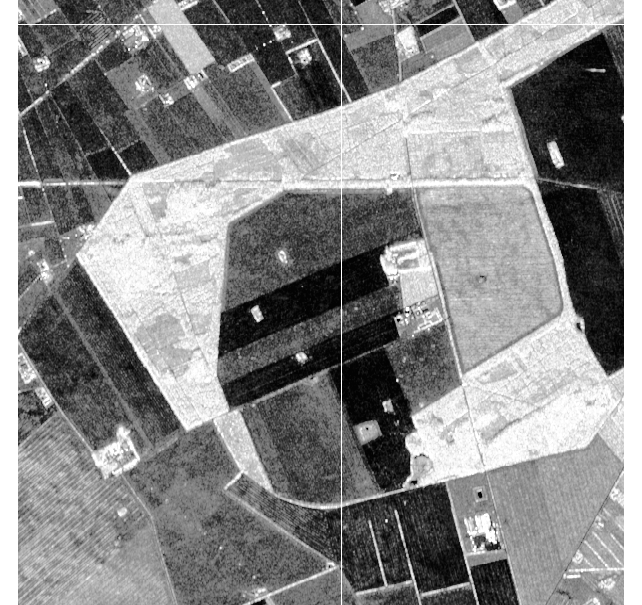}}
  \subfigure[Improved Sigma filter\label{fig:SAR-Foulum_LeeFilter}]{\includegraphics[width=.18\linewidth,viewport= 351 1 600 350,clip=TRUE]{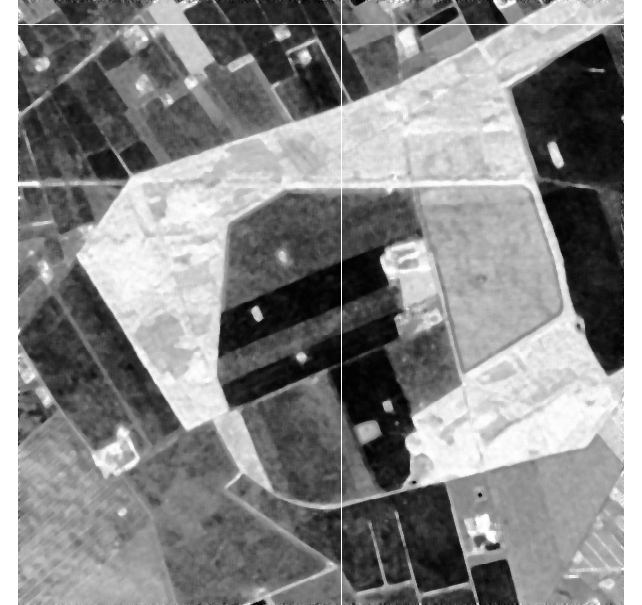}}
  \subfigure[SRAD filter\label{fig:SAR-Foulum_SRAD}]{\includegraphics[width=.18\linewidth,viewport= 351 1 600 350,clip=TRUE]{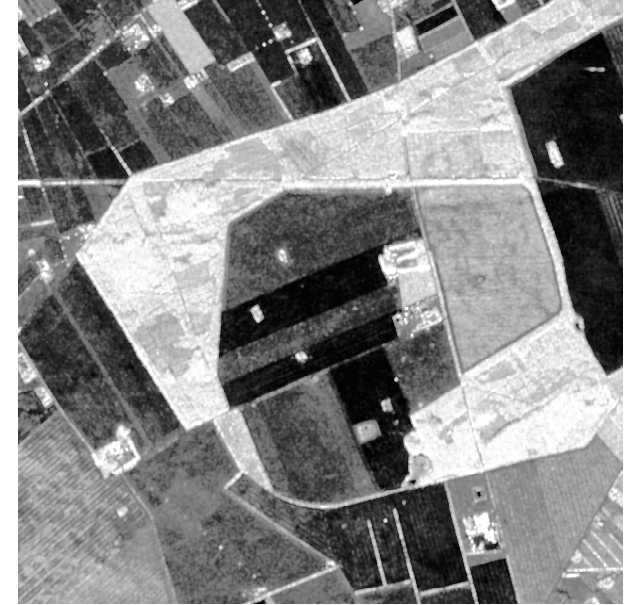}}
  \subfigure[SDNM filter\label{fig:SAR-Foulum_TorresFilter}]{\includegraphics[width=.18\linewidth,viewport= 351 1 600 350,clip=TRUE]{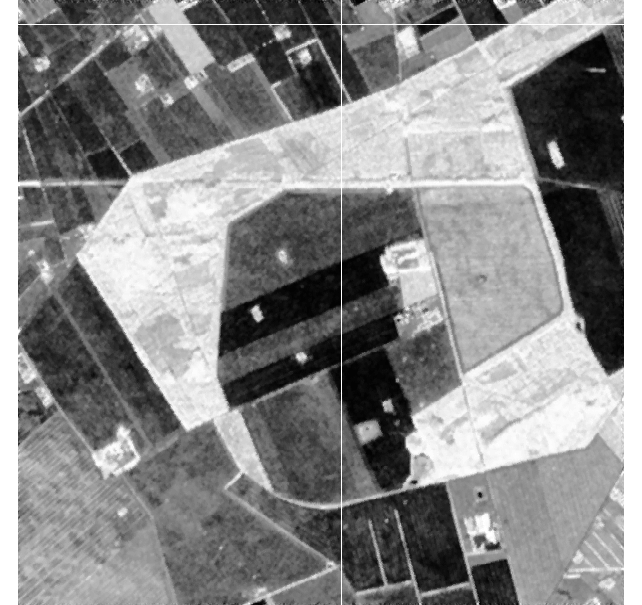}}
  \subfigure[SDNLM filter\label{fig:SAR-Foulum_ProposedFilter}]{\includegraphics[width=.18 	\linewidth,viewport= 351 1 600 350,clip=TRUE]{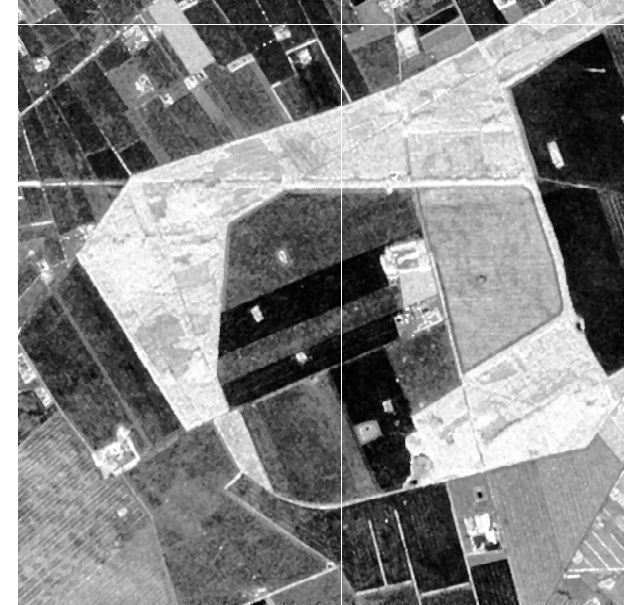}}
  \subfigure[$1$-D analysis on row $300$\label{fig:SAR-Foulum_Analysis}]{\includegraphics[width=.75\linewidth]{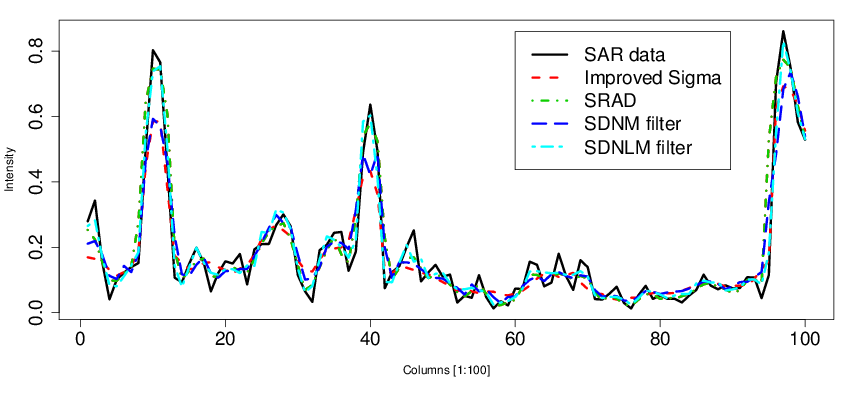}}
\caption{Real data, filtered versions and $1$-D analysis of row $300$.}
\label{fig:SARdataFoulum-Full}
\end{figure*}

Figure~\ref{fig:SARdataFoulum-Full} presents the results of applying the filters to an image obtained by the Danish EMISAR L-band fully polarimetric sensor over agricultural fields in Foulum, Denmark.
The original $250\times350$ pixels image of the HH intensity band is shown in Fig.~\ref{fig:SARdataFoulum}, its filtered versions by the Improved Sigma, SRAD, SDNM and SDNLM filters are presented in Figs.~\ref{fig:SAR-Foulum_LeeFilter}, \ref{fig:SAR-Foulum_SRAD}, \ref{fig:SAR-Foulum_TorresFilter} and~\ref{fig:SAR-Foulum_ProposedFilter}, respectively.
Figure~\ref{fig:SAR-Foulum_Analysis} presents the values of row $300$.

Table~\ref{tab:QIndexesRealImages} presents the observed $\operatorname{BRISQUE}$ index and equivalent number of looks on the real image.
It is noticeable that the proposed technique produces better results than the Improved Sigma, SRAD and SDNM filters regarding the $\operatorname{BRISQUE}$ index by a significant margin.
As expected, it provides a smaller increase in the equivalent number of looks, especially in the second image, which exhibits more spatial variability and thus leads to a more conservative filter.

\begin{table}[hbt]
\centering
\caption{Image quality indexes in the real SAR image.}
\begin{tabular}{l r@{.}l r@{.}l}
\toprule
\textbf{Filtered} & \multicolumn{2}{c}{$\operatorname{BRISQUE}$} & \multicolumn{2}{c}{\multirow{2}{*}{\textsf{ENL}}} \\
\textbf{Versions} & \multicolumn{2}{c}{\textbf{Index}} & \multicolumn{2}{c}{~} \\
\midrule
Improved Sigma & 49&239 & 4&581  \\
SRAD           & 39&235 & 3&101  \\
SDNM           & 39&164 & \textbf{4}&\textbf{661} \\
SDNLM          & \textbf{37}&\textbf{384} & 4&539 \\ 
\bottomrule 
\end{tabular}
\label{tab:QIndexesRealImages}
\end{table}

\section{Conclusions}\label{sec:conclu}

This paper presented an assessment of the filter based on stochastic distances for speckle noise reduction.
The proposal was compared with the Improved Sigma filter and other more sophisticated filters, using a protocol based on Monte Carlo experiences and real images. 
Moreover, the $\beta_{\rho}$, $Q$ index and $\operatorname{BRISQUE}$ model were used to assert the proposal.
The proposed filter outperforms the Improved Sigma filter in five out of six quality measures.
Other significance levels will be tested, along with different points of the parameter space of the simulation in order to have a more complete assessment of the proposal.

\bibliographystyle{IEEEtran}
\bibliography{ref_WTD_Sibgrapi2013}

\begin{thebibliography}{10}
\providecommand{\url}[1]{#1}
\csname url@samestyle\endcsname
\providecommand{\newblock}{\relax}
\providecommand{\bibinfo}[2]{#2}
\providecommand{\BIBentrySTDinterwordspacing}{\spaceskip=0pt\relax}
\providecommand{\BIBentryALTinterwordstretchfactor}{4}
\providecommand{\BIBentryALTinterwordspacing}{\spaceskip=\fontdimen2\font plus
\BIBentryALTinterwordstretchfactor\fontdimen3\font minus
  \fontdimen4\font\relax}
\providecommand{\BIBforeignlanguage}[2]{{%
\expandafter\ifx\csname l@#1\endcsname\relax
\typeout{** WARNING: IEEEtran.bst: No hyphenation pattern has been}%
\typeout{** loaded for the language `#1'. Using the pattern for}%
\typeout{** the default language instead.}%
\else
\language=\csname l@#1\endcsname
\fi
#2}}
\providecommand{\BIBdecl}{\relax}
\BIBdecl

\bibitem{Goodman1976}
J.~W. Goodman, ``Some fundamental properties of speckle,'' \emph{Journal of the
  Optical Society of America}, vol.~66, no.~11, pp. 1145--1150, 1976.

\bibitem{Lee1999PolSARspeckleFiltering}
J.-S. Lee, M.~R. Grunes, and G.~de~Grandi, ``{Polarimetric SAR} speckle
  filtering and its implication for classification,'' \emph{IEEE Transactions
  on Geoscience and Remote Sensing}, vol.~37, no.~5, pp. 2363--2373, 1999.

\bibitem{Lee1991MultiPol_SARimagery}
J.-S. Lee, M.~R. Grunes, and S.~A. Mango, ``Speckle reduction in
  multipolarization, multifrequency {SAR} imagery,'' \emph{IEEE Transactions on
  Geoscience and Remote Sensing}, vol.~29, no.~4, pp. 535--544, 1991.

\bibitem{Lee2006ScatteringModelBased_SpeckleFiltering}
J.-S. Lee, M.~R. Grunes, D.~L. Schuler, E.~Pottier, and L.~Ferro-Famil,
  ``Scattering-model-based speckle filtering of {Polarimetric SAR} data,''
  \emph{IEEE Transactions on Geoscience and Remote Sensing}, vol.~44, no.~1,
  pp. 176--187, 2006.

\bibitem{Lee2009}
J.-S. Lee, J.-H. Wen, T.~L. Ainsworth, K.-S. Chen, and A.~J. Chen, ``Improved
  sigma filter for speckle filtering of {SAR Imagery},'' \emph{IEEE
  Transactions on Geoscience and Remote Sensing}, vol.~47, no.~1, pp. 202--213,
  2009.

\bibitem{Cetin2001FeatureEnhancedSAR}
M.~{\c C}etin and W.~C. Karl, ``Feature-enhanced synthetic aperture radar image
  formation based on nonquadratic regularization,'' \emph{IEEE Transactions on
  Image Processing}, vol.~10, no.~4, pp. 623--631, 2001.

\bibitem{Osher2005IteratedRegularizationMethod}
S.~Osher, M.~Burger, D.~Goldfarb, J.~Xu, and W.~Yin, ``An iterated
  regularization method for total variation-based image restoration,''
  \emph{Multiscale Modeling \& Simulation}, vol.~4, no.~2, pp. 460--489, 2005.

\bibitem{Buades2005DenoisingAlgorithms}
A.~Buades, B.~Coll, and J.~M. Morel, ``\BIBforeignlanguage{{English}}{A review
  of image denoising algorithms, with a new one},''
  \emph{\BIBforeignlanguage{{English}}{Multiscale Modeling \& Simulation}},
  vol.~4, no.~2, pp. 490--530, 2005.

\bibitem{NonGaussianPatchSimilarity}
C.-A. Deledalle, F.~Tupin, and L.~Denis, ``Patch similarity under non gaussian
  noise,'' in \emph{IEEE International Conference on Image Processing (ICIP)},
  2011, pp. 1845--1848.

\bibitem{Deledalle2009IterativeDenoising}
C.-A. Deledalle, L.~Denis, and F.~Tupin, ``Iterative weighted maximum
  likelihood denoising with probabilistic patch-based weights,'' \emph{IEEE
  Transactions on Image Processing}, vol.~18, no.~2, pp. 2661--2672, 2009.

\bibitem{Deledalle2010NonlocalSARandInSAR}
C.-A. Deledalle, F.~Tupin, and L.~Denis, ``A non-local approach for {SAR} and
  interferometric {SAR} denoising,'' in \emph{IEEE International Geoscience and
  Remote Sensing Symposium (IGARSS)}, Honolulu, 2010, pp. 714--717.

\bibitem{Coupe2009NLM_FilteringUltrasound}
P.~Coup\'e, P.~Hellier, C.~Kervrann, and C.~Barillot, ``Nonlocal means-based
  speckle filtering for ultrasound images,'' \emph{IEEE Transactions Image
  Processing}, vol.~18, no.~10, pp. 2221--2229, 2009.

\bibitem{Kervrann2007BayesianNLM}
C.~Kervrann, J.~Boulanger, and P.~Coup\'e, ``Bayesian non-local means filter,
  image redundancy and adaptive dictionaries for noise removal,'' in
  \emph{Proc. Conf. Scale-Space and Variational Meth}, Ischia, 2007, pp.
  520--532.

\bibitem{Parrilli2012}
S.~Parrilli, M.~Poderico, C.~V. Angelino, and L.~Verdoliva, ``A nonlocal {SAR}
  image denoising algorithm based on llmmse wavelet shrinkage,'' \emph{IEEE
  Transactions on Geoscience and Remote Sensing}, vol.~50, no.~2, pp. 606--616,
  2012.

\bibitem{Dabov2007}
K.~Dabov, A.~Foi, V.~Katkovnik, and K.~Egiazarian, ``Image denoising by sparse
  3-d transform-domain collaborative filtering,'' \emph{IEEE Transactions on
  Image Processing}, vol.~18, no.~8, pp. 2080--2095, 2007.

\bibitem{Torres2012IGARSS}
L.~Torres, T.~Cavalcante, and A.~C. Frery, ``A new algorithm of speckle
  filtering using stochastic distances,'' in \emph{IEEE International
  Geoscience and Remote Sensing Symposium (IGARSS)}, Munich, 2012.

\bibitem{SpeckleReductionStochasticDistancesCIARP2012}
L.~Torres, T.~Cavalcante, and A.~Frery, ``Speckle reduction using stochastic
  distances,'' in \emph{Pattern Recognition, Image Analysis, Computer Vision,
  and Applications}, ser. Lecture Notes in Computer Science, {L. Alvarez et
  al.}, Ed., vol. 7441.\hskip 1em plus 0.5em minus 0.4em\relax Buenos Aires:
  Springer, 2012, pp. 632--639.

\bibitem{NagaoMatsuyama}
M.~Nagao and T.~Matsuyama, ``Edge preserving smoothing,'' \emph{Computer
  Graphics and Image Processing}, vol.~9, no.~4, pp. 394--407, 1979.

\bibitem{TorresSpeckleReductionPolSAR}
L.~Torres, S.~J.~S. {Sant'Anna}, C.~C. Freitas, and A.~C. Frery, ``Speckle
  reduction in polarimetric {SAR} imagery with stochastic distances and
  nonlocal means,'' \emph{Pattern Recognition}, April 2013, in press.

\bibitem{WMAPSpeckleFilterWavelet}
S.~S{\o}lbo and T.~Eltoft, ``{$\Gamma$-WMAP}: a statistical speckle filter
  operating in the wavelet domain,'' \emph{International Journal of Remote
  Sensing}, vol.~25, no.~5, pp. 1019--1036, 2004.

\bibitem{Nascimento2010}
A.~D.~C. Nascimento, R.~J. Cintra, and A.~C. Frery, ``Hypothesis testing in
  speckled data with stochastic distances,'' \emph{IEEE Transactions on
  Geoscience and Remote Sensing}, vol.~48, no.~1, pp. 373--385, 2010.

\bibitem{Salicru1994}
M.~Salicr{\'u}, D.~Morales, M.~L. Men{\'e}ndez, and L.~Pardo, ``On the
  applications of divergence type measures in testing statistical hypotheses,''
  \emph{Journal of Multivariate Analysis}, vol.~21, no.~2, pp. 372--391, 1994.

\bibitem{UIQIndex}
Z.~Wang and A.~C. Bovik, ``A universal image quality index,'' \emph{IEEE Signal
  Processing Letters}, vol.~9, no.~3, pp. 81--84, 2002.

\bibitem{Moschetti2006}
E.~Moschetti, M.~G. Palacio, M.~Picco, O.~H. Bustos, and A.~C. Frery, ``On the
  use of {L}ee's protocol for speckle-reducing techniques,'' \emph{Latin
  American Applied Research}, vol.~36, no.~2, pp. 115--121, 2006.

\bibitem{Mittal2012NoReferenceIQASpatialDomain}
A.~Mittal, A.~K. Moorthy, and A.~C. Bovik, ``No-reference image quality
  assessment in the spatial domain,'' \emph{IEEE Transactions on Image
  Processing}, vol.~21, no.~12, pp. 4695--4708, 2012.

\bibitem{Yu2002SpeckleReducingAnisotropicDiffusion}
Y.~Yu and S.~Acton, ``Speckle reducing anisotropic diffusion,'' \emph{IEEE
  Transactions on Image Processing}, vol.~11, no.~11, pp. 1260--1270, 2002.

\end{thebibliography}

\end{document}